\documentclass[prc,aps,amsfonts,amsmath,superscriptaddress,floatfix,twocolumn]{revtex4}
\usepackage{graphicx,float} 
\usepackage{epsfig} 
\usepackage{rotating}
\usepackage{mwe}    
\usepackage{subfig}
\usepackage{braket}

\begin{document}

\title{Use of quantality in nuclei and many-body systems}

\author{J.-P. Ebran}
\affiliation{CEA, DAM, DIF, 91297 Arpajon, France}
\affiliation{Universit\'e Paris-Saclay, CEA, Laboratoire Mati\`ere en Conditions Extr\^emes, 91680 Bruy\`eres-le-Ch\^atel, France}
\author{L. Heitz}
\affiliation{IJCLab, Universit\'e Paris-Saclay IN2P3-CNRS, F-91406 Orsay Cedex, France}
\author{E. Khan}
\affiliation{IJCLab, Universit\'e Paris-Saclay IN2P3-CNRS, F-91406 Orsay Cedex, France}
\affiliation{Institut Universitaire de France (IUF)}

\begin{abstract} 

The use of quantality is discussed in the case of nuclei and other many-body systems such as atomic electrons. This dimensionless quantity is known
to indicate when a many-body system behaves like a crystal or a quantum liquid. Its role is further analyzed by showing its relation to the scattering length. The emergence of a fundamental lengthscale, the limit radius, is also shown. It corresponds to the hard-core of the nucleon-nucleon interaction in the case of nucleons, and to a value close to the Bohr radius in the case of atomic electrons. 
The occurrence of a cluster phase in nuclei is analyzed using the quantality through its relation to the localization parameter, allowing for the identification of both the number of nucleons and the density as control parameters for the occurrence of this phase. 
The relation of the quantality to the magnitude of the interaction also exhibits a third dimensionless parameter, monitoring the magnitude of the spin-orbit effect in finite systems, through the realization of the pseudo-spin symmetry.
The impact of quantality on the spin-orbit effect is compared in various many-body systems. The role of quantality in the relative effect of the binding energy and the shell one is also analyzed in nuclei. Finally, additional
dimensionless quantities are proposed from the generalization of the quantality.
Nuclei are found to be exceptional systems because all their dimensionless quantities are close to the order of unity, at variance with other many-body systems.
\end{abstract}
 


\date{\today}

\maketitle

\section{Introduction}

 Quantality, a powerful dimensionless ratio first introduced by de Boer \cite{boe48}, was used by Mottelson to indicate when a system behaves like a quantum liquid  (QL) rather than a
crystal. It is larger for quantum liquid states than for crystal ones \cite{mot96,mot99}. The former have constituents with delocalized wave-functions, leading to a homogeneous density, whereas the latter have localized constituents at given nodes. Superfluid Helium systems, nuclei, and atomic electrons are of QL type \cite{mot96,zin08,zin13,cla24}.  More precisely, nuclei and atomic electrons are found to be the two most QL-like systems, having larger values of the quantality than superfluid Helium systems.

Let us consider a many-body system composed of constituents of mass m and interacting with an interaction of typical
magnitude V$_0$. The inter-constituent distance at equilibrium is r$_0$. These 3 quantities characterize the main behavior of the system as they correspond to the input of its equation of motion. Instead of solving it as exactly as possible, a complementary way to predict the qualitative behavior of the system is to use indicators, incorporating step by step
quantum mechanical effects. As a first step, the quantality is defined as the ratio of the Zero Point Energy (ZPE) T$_0$ to the magnitude of the potential V$_0$:

\begin{equation}
\Lambda \equiv
\frac{2T_0}{V_0}=\frac{\hbar^2}{mr_0^2V_0}
\label{la}
\end{equation}

Here, the ZPE T$_0$ is defined as the kinetic energy due to the equilibrium distance r$_0$, through the Heisenberg relation:

\begin{equation}
T_0=\frac{\hbar^2}{2mr_0^2}
\label{t0}
\end{equation}

Quantality was used by Mottelson to characterize QL vs crystal behavior of systems \cite{mot96,mot99}. The typical value calculated with Eq. (\ref{la})
is $\Lambda \simeq$ 10$^{-2,-3}$ in the case of crystals like atoms
and molecules, and $\Lambda \simeq$ 0.1-1 (i.e. more than one order of magnitude larger) in the case of QL such as
$^4$He, nuclei or electrons in atoms (see Table I). In the case of
nuclei, $\Lambda \simeq$ 0.4, using typical values of the nucleon-nucleon interaction, namely V$_0$ $\simeq$ 100 MeV and r$_0$
$\simeq$ 1 fm \cite{mot96,mot99}. The quantality indicates the importance to analyze the relation between the ZPE and the potential energy in finite systems. Moreover,
among the quantality values of Table I, nuclei and atomic electrons have the largest one, indicating strong quantum effects. 
It should be noted that i) quantality is valid for both fermions and bosons,
ii) the  quantum liquid nature of electron in atoms has been discussed in Ref. \cite{mot96},
and iii) results have to be considered within one order of magnitude, due to the semi-quantitative nature of the approach.

\begin{table*}[t]
\renewcommand{\arraystretch}{1.5}
\centering
\begin{tabular}{ccccccccccccc}
\hline \hline
        Constituent  & m & r$_0$ (nm) & V$_0$ (eV) & T$_0$ (eV)& $\Lambda$ &State 
	& $\alpha$ & $\eta$ & $\cal{A}$  &   ZPEv  & QV & TI \\
\hline
$^{20}$Ne atom & 20 & 0.31 & 31 10$^{-4}$ & 1.1 10$^{-5}$ & 0.007 & crystal  &4.9 10$^{-6}$ & -6.1 10$^{12}$&12.0& 3.3 10$^{-8}$  & 5.5 10$^{-21}$ & 8.0 10$^{-19}$  \\
H$_2$ molecule & 2 & 0.33  & 32 10$^{-4}$ & 9.5 10$^{-5}$ & 0.06 & crystal &5.3 10$^{-6}$& -5.9 10$^{11}$& 4.1   & 3.2 10$^{-7}$ &  5.3 10$^{-19}$  & 9.0 10$^{-18}$  \\ 
$^4$He atom & 4 & 0.29  & 8.6 10$^{-4}$ & 6.0 10$^{-5}$ & 0.14 & QL & 1.2 10$^{-6}$ & -4.4 10$^{12}$& 2.7   &  1.9 10$^{-7}$  & 4.7 10$^{-20}$   & 2.7 10$^{-19}$ \\
$^3$He atom & 3 & 0.29  & 8.6 10$^{-4}$ & 0.8 10$^{-5}$  & 0.19 & QL  &1.2 10$^{-6}$ & -3.2 10$^{12}$ & 2.3   &   2.6 10$^{-7}$  & 7.7 10$^{-20}$  & 3.8 10$^{-19}$ \\
Nucleon & 1 & 10$^{-6}$ & 100 10$^{6}$ & 25 10$^{6}$ & 0.41 & QL  & 0.51 & 1.3 & 1.6   &   0.18    & 0.02  & 0.04 \\
e$^-$ in atoms & 5 10$^{-4}$ & 0.05 & 10 & 16 & 3.1 & QL & 2.5 10$^{-3}$& -5 10$^{4}$&0.6  & 8.1 10$^{-3}$   &  1.6 10$^{-7}$   & 5.0 10$^{-8}$ \\
\hline \hline
\end{tabular}
\caption{Constituent mass m, interaction lengthscale r$_0$, its magnitude V$_0$, zero point kinetic energy T$_0$, quantality $\Lambda$ \cite{mot96},
effective coupling constant $\alpha$, spin-orbit parameter $\eta$ \cite{ebr16} and action $\cal{A}$ 
for various many-body systems.  m is given in units of a nucleon mass for convenience. The Zero Point Energy velocity (ZPEv), quadratic velocity (QV) and total interaction (TI) are introduced in section V.}
\label{tab:quanta}
\end{table*}

An intermediate phase could appear between the crystal and the quantum liquid phase, the clusterized one \cite{gre02,fal05,yan07}. The introduction of another dimensionless quantity, the localization parameter \cite{ebr12}, could help, for instance, to understand why light nuclei are more clustered than heavy ones \cite{fre18}. This parameter depends on the spatial dispersion of the nucleonic wavefunction which is determined by the equation of motion. Therefore, the localization parameter is richer than the quantality because it takes into account finite-size effects. It is analogous to the Wigner or Br\"uckner parameter introduced in plasma physics or in condensed matter physics \cite{wig34}. It allows for a unified understanding of cluster phases as transitional states between crystal and quantum liquid phases, also in nuclei \cite{ebr18,ebr17,ebr14a,fal05}.  The relation between the quantality and the localization parameter was studied in \cite{ebr13}.

Therefore, understanding $\alpha$ clusterization in nuclei as a different phase, compared to the QL one is relevant. 
The use of the localization parameter calculated both analytically with the Harmonic Oscillator (HO) approximation, and with covariant Energy Density Functionnals (EDF), allowed to understand nuclear clusterization 
as a nuclear transitional phase, compared to the usual nuclear QL one \cite{ebr12,ebr13}. A transition from a gas of  $\alpha$ particles to nuclear liquid was also related to the properties of the nucleon-nucleon interaction with the lattice calculations \cite{elh16,fre18}. More recently, Effective Field Theories (EFTs)  also tackle the occurrence of cluster phase in nuclei \cite{daw20}. Recent Ab initio PGCM calculations also confirm clusteristation traces in light nuclei, especially in $^{16}$O \cite{fro22}. 

Control parameters allowing for the appearance of the cluster phase in nuclei have been discussed through these models, providing a broader understanding of the occurrence of clusterization than the only vicinity of the multi-alpha particle threshold provided by the seminal Ikeda conjecture \cite{ike68}: the nucleon-nucleon interaction \cite{elh16}, the depth of the confining potential \cite{ebr12}, the nucleon number \cite{ebr13,ebr14a}, the nuclear deformation \cite{ebr14,ebr14a}, the nuclear density \cite{rop10,typ14,sch13,ebr20}, or the temperature \cite{yuk22} could drive the transition from a nuclear quantum liquid state to a clusterized one. 
These studies allow to consider nuclear clusterization as an intermediate phase, as in other systems \cite{gre02,fal05,yan07}, which can occur in nuclear ground states.  On the experimental side, recent signals using alpha knock-out reactions \cite{tan21,li23} attracted a lot of attention as they allow to probe the degree of clusterization over the nuclear chart, from light nuclei (Be) to heavy ones such as the Tin isotopes.

Despite its powerfulness, the role of quantality has been very scarcely studied in nuclear physics \cite{mot96,mot99,zin13,ebr14a,cla24}. In this work, we further explore how quantality can be used to understand the main differences between atomic electrons and nucleons in their quantum liquid phases. It is then used to identify the control parameters for the cluster phase in nuclei, namely the role of the number of nucleons and the density.  The relation to the low-density transition towards clusterization is analyzed. The use of quantality, as a dimensionless ratio, to understand its relation with the interaction coupling constants leads to a fundamental relation, allowing to study the magnitude of the spin-orbit effect in finite nuclei. The relative magnitude of the binding energy to the shell-structure one is also shown to depend on the quantality. 

The present work aims to analyze the behavior of many-body systems with simple but powerful indicators, by progressively refining their quantum description. First, 
the role of the ZPE is discussed. Then, the inclusion of the dispersion of the wave-function of the constituents, from the solution of a e.g. Schr\"odinger equation is considered. As a third step, the relation with the spin-orbit effect is given, based on the non-relativistic reduction of the Dirac equation. It will be shown how quantality plays a pivotal role during all these three stages. In section II, the study of the occurence of bound states from a given interaction first provides a relation between the quantality and the scattering length. Then, the emergence of a fundamental lengthscale, the limit radius, is studied in relation to the ZPE and the quantality. The relation between the quantality and the localization parameter is discussed in section III, showing how the appearance of clusters in the low-density regime is recovered. Section IV relates the quantality to the dimensionless magnitude of the interaction, showing how the spin-orbit effect in finite systems can be deduced. Finally, a systematic study of possible dimensionless quantities is provided in Section V, generalizing the concept of quantality.

\section{Quantality and bound states}

\subsection{General definition of quantality}

In a more general manner, the quantality could be defined as the ratio of the mean values of the kinetic energy to the potential one:

\begin{equation}
\Lambda =
\frac{2\langle T \rangle}{\langle V \rangle}
\label{lagene}
\end{equation}

where Eq. (\ref{la}) is the application to the phenomenological case with a given potential and kinetic energy values. Eq. (\ref{lagene}) allows the calculation of the value of the quantality for any interaction and many-body method, by using the corresponding wave functions and operators. Such a control parameter might prove useful in Ab Initio approaches. For instance, it could be possible to monitor the degree of clusterisation as a function of the similarity renormalisation group evolution of the bare chiral potential \cite{fur12,bog07,her20},  or of the many-body scheme at stake \cite{lee09,hag10,fro22}. One can then study with these approaches if, e.g., the quantum liquid behavior of the system is in agreement with the calculated $\Lambda$ value. In the present work, we will keep the use of Eq. (\ref{la}), as it already paves the way for 
numerous developments.

\subsection{Quantality and the scattering length}

A first way to analyze the occurence of bound states from a given interaction is to relate the quantality to the scattering length. Let us consider an inter-constituent distance r$_0$ and describe the two-body interaction by a spherically symmetric finite well. One can then derive an analytical expression of the scattering length \cite{mac23}. Using the present notations and the definition (\ref{la}) of the quantality, the scattering length becomes:

\begin{equation}
a=r_0\left(1-\sqrt{\Lambda}\tan\left(\frac{1}{\sqrt{\Lambda}}\right)\right)
\label{la72}
\end{equation}

\begin{figure}[tb]
\begin{center}
\scalebox{0.35}{\includegraphics{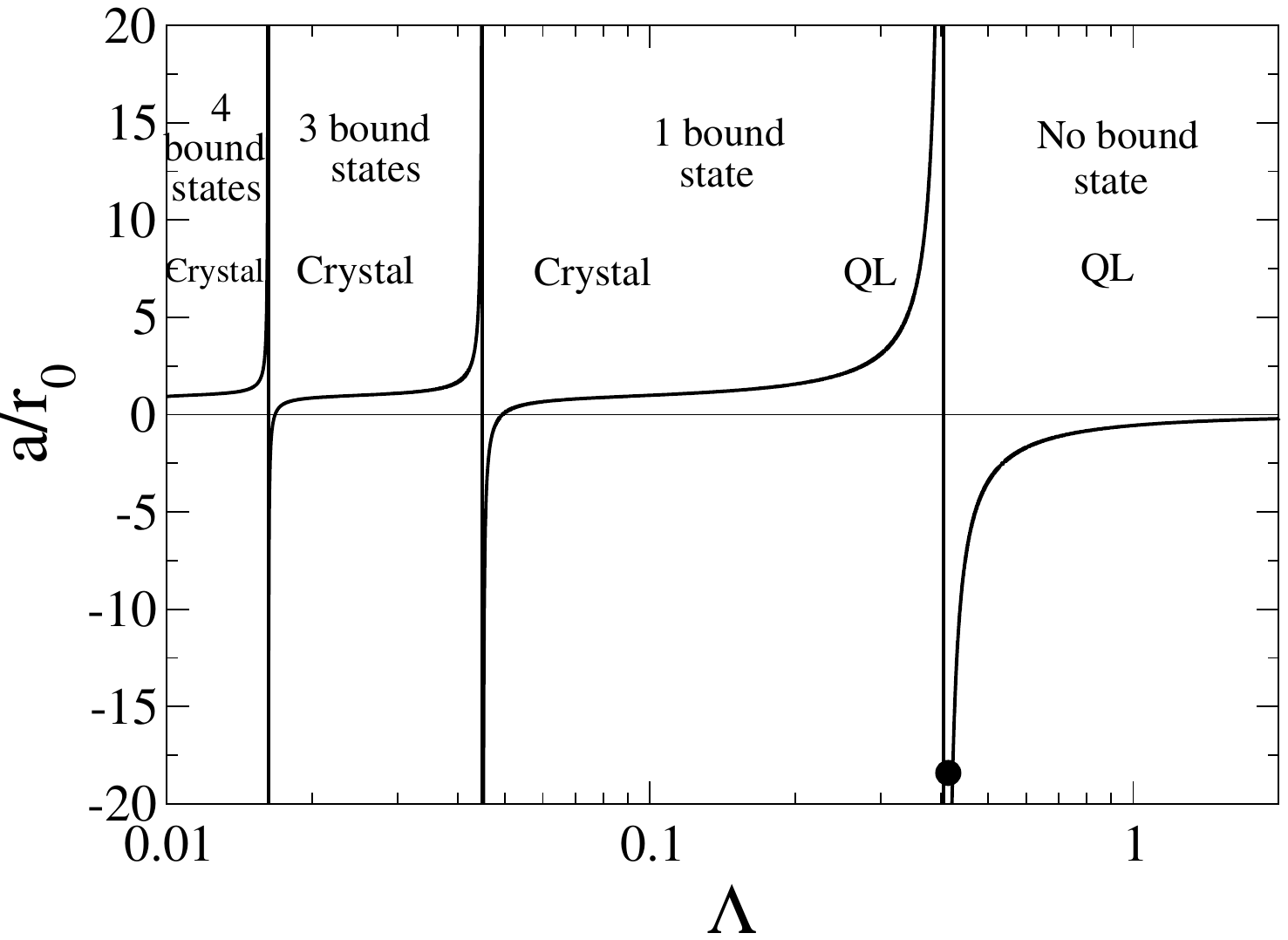}}
 \caption{Scattering length a over the inter-constituent distance r$_0$ as a function of the quantality $\Lambda$ using Eq. (\ref{la72}). The quantum liquid (QL) or crystal behavior is obtained from the values of the quantality (see Table \ref{tab:quanta}). The black dot corresponds to the case of the nucleon-nucleon interaction.}    
\label{fig:scl}
\end{center}
\end{figure}

The behavior of a/r$_0$ as a function of the quantality is shown in Fig. \ref{fig:scl}. The unitarity limit, for which the scattering length diverges, is obtained for specific values of $\Lambda$, namely from Eq. (\ref{la72}):

\begin{equation}
\Lambda=\frac{4}{\pi^2(2n+1)^2}
\end{equation}

where n=0,1,2, ... The first unitarity limit corresponds to $\Lambda\simeq$0.4 and the second one for $\Lambda\simeq$0.05. The former value of the quantality is in the QL regime, whereas the latter is in the crystal one (see Table \ref{tab:quanta}). In the case of the nucleon-nucleon interaction, it is striking that the corresponding quantality value of  $\Lambda\simeq$0.4 allows
to recover a scattering length close to the unitarity limit, in agreement with the phenomenology (a$\simeq$ -19 fm in the case of the neutron-neutron interaction in the s-wave \cite{bm69}). It should be noted that recent Ab initio calculations
analyzed how clusterization emerges at the unitarity limit \cite{daw20}. 
 
When the quantality decreases, typically for $\Lambda<$0.1, the system is expected to behave as a crystal (see Table \ref{tab:quanta}). Concomitantly, additional
bound states become possible, as seen on  Fig. \ref{fig:scl}. Therefore, two constituents in crystal systems can have more bound states than in QL.

\subsection{Quantality and the limit radius}

The inter-constituent distance r$_0$, has also a lower boundary r$_l$: below this limit radius r$_l$, the ZPE is large enough to overcome the potential energy and make the system unbound. The corresponding limit condition T$_0$=V$_0$ leads to

\begin{equation}
r_l=\frac{\hbar}{\sqrt{2mV_0}}
\label{rl}
\end{equation}

\begin{figure}[tb]
\begin{center}
\scalebox{0.70}{\includegraphics{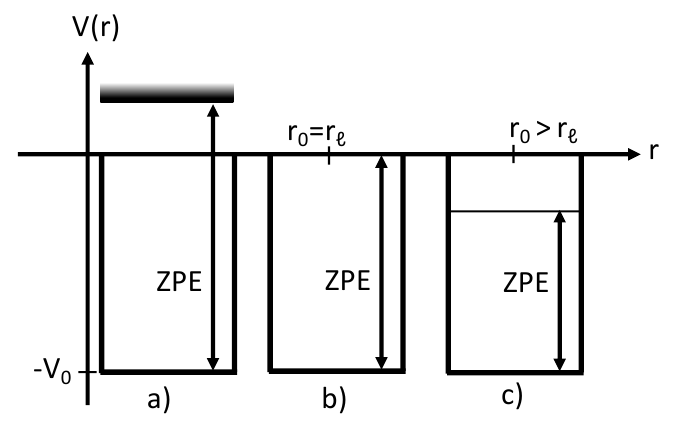}}
 \caption{Schematic energy position of a 2 constituents system in the case of the spherically symmetric finite well approximation of the inter-constituent interaction for  a) r$_0<$r$_l$, b) r$_0=$r$_l$ and c) r$_0>$r$_l$.  }    
\label{fig:zpepot}
\end{center}
\end{figure}

Fig. \ref{fig:zpepot} gives a view of the interplay of V$_0$, r$_0$, r$_l$ and the ZPE considering the spherically symmetric finite well approximation of the inter-constituent interaction, in the three following cases: r$_0<$r$_l$, r$_0=$r$_l$ and r$_0>$r$_l$, for which the system is unbound, at the boundary limit and bound, respectively.

The quantality (\ref{la}) can then be written as:

\begin{equation}
\Lambda =2\left(\frac{r_l}{r_0}\right)^2
\label{la2}
\end{equation}

Close to the limit case, such as for electrons in atoms, r$_0$=r$_l$ and the quantality takes its maximal value, namely $\Lambda\simeq$2. This is confirmed by the phenomenological value of quantality \cite{mot96,mot99},
reproduced in Table I. In other systems, $\Lambda\le$2. For instance, in the case of nuclei, $\Lambda\simeq$1. This is a specific feature of nuclei: $\Lambda$ is of the order of unity, as well as their
other dimensionless quantities. We define it as specificness and its origin shall be discussed in section \ref{subkey}.
For instance, although both atomic electrons and nuclei have a quantality value close to unity, it will not be the case for other dimensionless quantities in the case of atomic electrons, contrarily to nuclei, as discussed in section IV. Eq (\ref{la}) also shows that $\Lambda$ monitors the delocalization effects with respect to the minimal radius r$_l$: the larger the equilibrium distance
r$_0$, the smaller the quantality with respect to the maximal value $\Lambda\simeq$2. 

Whether the particle is at equilibrium at a distance r$_0$ equal to or larger than r$_l$ may depend on several causes, such as delocalization effects.
In nuclei, using the typical magnitude V$_0$ of the nucleon-nucleon interaction gives r$_l$=0.5 fm from Eq. (\ref{rl}). This corresponds to the size of the repulsive core of the interaction. It can be understood with the specificness of nuclear systems: as an order of magnitude reasoning, V$_0$r$_l\simeq\hbar$c. Put in other words, the r$_l$=0.5 fm short-scale distance corresponds to an energy of $\hbar$c/$r_l$ $\simeq$ 400 MeV which defines an energy cutoff above which the ZPE is too large for the system to be bound. Interestingly, this about 400 MeV value also corresponds to the phenomenological determination of the cutoff parameter in chiral EFT \cite{mar13,mac24}, pointing to a possible link with the ZPE limit, which remains to be investigated.  

Eq (\ref{la2}) can be rewritten as

\begin{equation}
r_0=r_l\sqrt{\frac{2}{\Lambda}}
\label{la3}
\end{equation}

The equilibrium distance r$_0$ remains larger than r$_l$ because $\Lambda\le$2. However, the smaller $\Lambda$, the larger inter-constituent distance r$_0$, and the smaller the delocalization effect, because the constituents are further apart. Therefore, $\Lambda$ drives localization effect. However, it misses finite-size effects. This will be taken into account in section III. 

In nuclei the limit radius, being of the order of the hard-core one, gives a straightforward reason why the hard-core is washed out: it corresponds to the minimal radius allowed by the ZPE, and in practice, the mean inter-nucleon distance is r$_0\simeq$ 1.2 fm, which is significantly larger. Therefore, due to a delocalization effect monitored by the quantality, the saturation density is lower than
the packing one, which would correspond to a nucleonic system with inter-constituent distance r$_l$, the size of the repulsive core of the interaction. As discussed by Mottelson, this is an important question to understand  \cite{bm69,fuk20}. More precisely, using the relation between the inter-constituent distance and the density:

 \begin{equation}
 r_{0,l} = \left(\frac{3}{4\pi\rho_{0,l}}\right)^\frac{1}{3},
 \label{la5}
 \end{equation}

Eq (\ref{la3}) leads to

\begin{equation}
\rho_l=\left(\frac{2}{\Lambda}\right)^{3/2}\rho_0\simeq 10\rho_0
\label{la4}
\end{equation}

where the quantality value $\Lambda$$\simeq$0.4 for nuclei has been used. Therefore, the packing density in nuclei is about one order magnitude larger than the saturation one ($\rho_0\simeq$0.16 fm$^{-3}$), which can be understood from the value of quantality: $\rho_l\simeq$1.5 fm$^{-3}$.

Turning to the case of electrons in atoms, and using the Coulomb energy for V$_0$:

 \begin{equation}
 V_0=\frac{e^2}{a_B},
 \end{equation}
 
 where  a$_B\equiv\hbar^2$/me$^2$ is  the Bohr radius, leads to r$_l$=a$_B$/$\sqrt{2}$, using Eq. (\ref{rl}). Since the equilibrium distance is r$_0$=a$_B$, this means that r$_0$ is close to its minimal value and that the ZPE is close to its maximal value: r$_0$=a$_B$=r$_l$$\sqrt{2}$. The present approach allows to realize that atomic electrons are an exceptional system because it is close to a limit case. Its origin can be found in the specific 1/r behavior of the Coulomb potential, interplaying with the 
1/r$^2$ behavior of the ZPE: equilibrium position involves spatial derivatives which are related, as the derivative of the potential energy can be expressed from the kinetic energy \cite{coh}. 

Therefore, the limit radius r$_l$ is a fundamental quantity, recovering both the hard-core size in nuclei as well as a value close to the Bohr radius in the case of atomic electrons. 

\section{Quantality and the cluster phase in nuclei}

A step further in including quantum effects in quantities monitoring the behavior of the system can be achieved by the use of the
localization parameter \cite{ebr12,ebr13}. It is defined as

 \begin{equation}
 \alpha_{loc}(r_0)\equiv \frac{\lambda}{r_0}  
 \label{defloc}
 \end{equation}

 where $\lambda$ is the typical spatial extension of the constituent wavefunction and r$_0$  the inter-constituent distance at equilibrium. 
 In order to study cluster phases, this dimensionless quantity is $>$1 for homogeneous quantum liquid phase, $<$1 in the crystal case and
 $\sim$1 in the transitional cluster case. This quantity is related to the so-called 
 Br\"uckner one in condensed matter (see Appendix A). The interest of the localization parameter is that it takes into account finite-size effects, compared to quantality, by introducing $\lambda$ as an additional quantity to consider in addition to m, r$_0$ and V$_0$.
 
 In the case of nuclei, the spatial extension of the nucleonic wave function $\lambda$  can be calculated microscopically from e.g. EDF approaches \cite{ebr18}. It can also be calculated from ab-initio or lattice EFT approaches for instance, by computing the spatial dispersion $\Delta$r of the corresponding wave functions. Similarly, the inter-constituent distance could be computed from lattice-EFT calculations, in order to obtain a microscopic value. Here, in order to analyze the main impacts of the localization parameter, we use the sound approximation by the confining length of a harmonic oscillator (HO) potential, which parameters are chosen to match the radius $R$ of a given nucleus \cite{ebr18}:

\begin{equation}
\lambda \simeq \sqrt{\frac{\hbar}{m\omega}}=\frac{\sqrt{\hbar R}}{(2mV_0)^\frac{1}{4}}
\label{disp}
\end{equation}

with $\hbar\omega$ the typical energy of the HO, $m$ the nucleon mass
and $V_0$ the depth of the confining potential. The typical confining potential V$_0$ has been taken as the one of the magnitude of the nucleon-nucleon interaction of Eq. (\ref{la}). 
This is well justified in nuclei, as the empirical values are 80 MeV \cite{ebr12} and 100 MeV \cite{mot96} for the depth of the confining potential, and the magnitude of the nucleon-nucleon interaction, respectively. A more detailed justification for this approximation is given in the appendix of Ref. \cite{ebr14}.

Using Eqs (\ref{rl}) and (\ref{la5}), Eq. (\ref{disp}) can be rewritten as 
\begin{equation}
\lambda \simeq \sqrt{r_lR}=\sqrt{r_0R}\left(\frac{\Lambda}{2}\right)^{1/4}
\label{displa}
\end{equation}

Eq (\ref{displa}) shows that the typical nucleonic dispersion can be interpreted as the
geometrical average of the limit radius from the ZPE and the nuclei radius. It also emphasizes the contribution of both the quantality and finite-size effects on the dispersion. It should be noted that this equation can also be applied to atomic electrons, the HO confining potential being a first-order approximation to any bound system. Since both r$_l$ and R are close to the Bohr radius, as discussed in the previous section, the dispersion of the electron wave function $\lambda$ is also close to the Bohr radius.

The localization parameter can be related to the quantality. Especially in nuclei, where R=r$_0$A$^{1/3}$, and using Eqs. (\ref{defloc}),(\ref{displa}):
\begin{equation}
\alpha_{loc} \simeq\left(\frac{\Lambda}{2}\right)^{1/4}A^{1/6}\simeq 0.7 A^{1/6},
\label{locf}
\end{equation}

showing the A dependence introduced by the localization parameter, compared to the quantality. This expression has been studied in previous works \cite{ebr14a,ebr18}, and allows to describe the more frequent cluster occurrence in light nuclei, compared to heavy ones. A value smaller or larger than 1 corresponds to localized or delocalized states in quantal systems, respectively. It should be noted that Eq (\ref{locf}) has also been confirmed by the use of microscopic calculations of the spatial dispersion of the wave function \cite{ebr13,ebr18}.

Having identified the number of nucleons as a control parameter for the occurrence of a cluster phase, the role of the density in such phases can now be analyzed with the present approach. 
The inter-constituent distance is related to the density of the system. Let us consider a many-body system where the inter-constituent distance can change compared to its equilibrium value, due for instance, to external conditions such as compression or dilatation of the system, as it can occur in heavy-ion collisions. It can also be that the density of the system itself is spatially dependent and hence have a lower density on the surface. For instance, finite nuclei display large fluctuations around $\rho_\text{sat}$, such that nucleons evolving in the low-density parts of finite nuclei could self-arrange into alpha-clusters \cite{rop10,typ14,sch13,ebr20},  as probed by recent (p,p$\alpha$) reactions \cite{tan21}. 

As an in-medium monitor for cluster phase occurrence where finite-size effect can be important, such as in nuclei or for atomic electrons, the localization parameter (\ref{defloc}) can be generalized to any inter-constituent distance $\bar{r}$:

\begin{equation}
 \alpha_{loc}=\frac{\lambda}{\bar{r}}  
 \label{locdef2}
 \end{equation}
 
 The mean interparticle-distance is related to the density of the system by 
 
 \begin{equation}
 \bar{r} = \left(\frac{3}{4\pi\rho}\right)^\frac{1}{3}. 
 \label{minr}
 \end{equation}
 In the case of nuclei, the equilibrium distance $\bar{r}$=r$_0$ corresponds to the saturation density $\rho_0$. 
 
Using R=$\bar{r}$A$^{1/3}$ owing to the short range of the nucleon-nucleon interaction and Eqs (\ref{disp}), (\ref{locdef2}), (\ref{minr}) lead to:

\begin{equation}
\alpha_{loc} = \sqrt{\frac{r_l}{\bar{r}}}A^{1/6}=\left(\frac{\rho}{\rho_l}A\right)^{1/6}
\label{eq:al}
\end{equation}
where $\rho_l$ is the density corresponding to the minimal radius. In nuclei, this is the packing density \cite{bm69,fuk20} because r$_l$ corresponds to the hard-core size, as discussed in section II.

\begin{figure}[tb]
\begin{center}
\scalebox{0.60}{\includegraphics{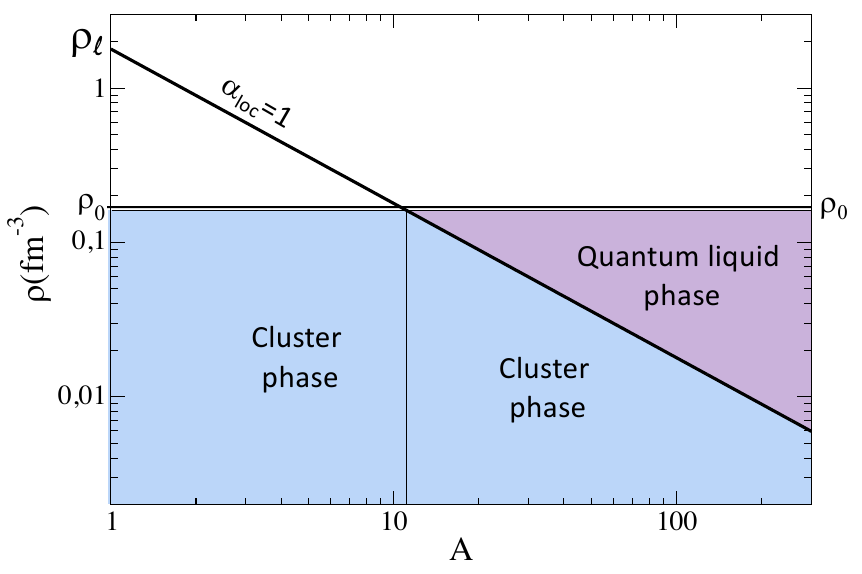}}
 \caption{Phase diagram of the cluster and quantum liquid states in nuclei, based on relation (\ref{mas}), showing the density and number nucleons as control parameters.}    
\label{fig:phaseCP}
\end{center}
\end{figure}

Clusters shall occur when the nucleon spatial extension is of the order the inter-nucleon distance, namely 
$\alpha_{loc}\lesssim$1 \cite{ebr12,ebr13,ebr14a}. In this case, Eq (\ref{eq:al}) brings a condition for cluster states:

\begin{equation}
\rho A\lesssim\rho_l=\left(\frac{2}{\Lambda}\right)^{3/2}\rho_0\simeq 10\rho_0,
\label{mas}
\end{equation}

using Eq. (\ref{la4}). Hence, cluster states can occur in light nuclei (A$\simeq$10) at saturation density or at a lower density than the saturation one in heavy nuclei (A$>$10): either the density or the number of nucleons can be a control parameter for the occurrence of cluster phase from a nuclear quantum liquid one. Interestingly, the threshold number of A$\sim$ 10 for clusterization is driven by the quantality. Eq. (\ref{mas}) also shows that in $^{120}$Sn, clusterization could
occur below a density of about one-tenth of the saturation density, that is on its very surface. 
These results are in agreement with nuclear phenomenology, where lighter nuclei and/or smaller density (the so-called Mott transition \cite{ste95,rop98,bey00,tak04}) trigger cluster states. This is
also in qualitative agreement with predictions of more microscopic approaches based e.g. on EDF, with or without explicit inclusion of alpha degrees of freedom \cite{sch13,typ14,ebr20}, illustrating the relevance of the present discussion. Fig. \ref{fig:phaseCP} summarizes the conditions for the cluster and quantum liquid phases induced by 
the condition (\ref{mas}), exhibiting both the density and the number of nucleons as control parameters. This picture, validated in the relativistic EDF, could be further investigated in other approaches such as ab-initio and EFT ones. 

Another way to trigger the cluster phase is to dilute the whole nucleus \cite{sch13,ebr20}.
In the case of the transition from a quantum liquid to an alpha clusterised state by a decrease of the density,
a more precise value of the density below which
all nucleons are likely to combine into alpha-particles and condensate can be given. Because of Pauli blocking effects, 
alpha-particles in the nuclear medium shall dissolve as they start overlapping with each others. The critical  Mott density
at which this happens turns out to be given by the dimensionless parameter $\alpha_{loc}$ where the ratio is now between 
the size of an alpha-particle $R_\alpha \simeq r_0A_\alpha^\frac{1}{3}$
and the mean internucleon distance $\bar{r}$. The Mott density is then solution
of $\alpha_{loc}=1$, that is, using Eq. (\ref{minr})

\begin{equation}
\frac{\rho^\alpha_\text{Mott}}{\rho_0}=\frac{1}{A_\alpha}\sim 0.25
\end{equation} 

Therefore, in the case of a global dilution of the nucleus, below about one fourth of
the saturation density, quartetting correlations are no more suppressed by Pauli blocking effects
and a full alpha-clustering should correspond to the favoured arrangement of alpha-conjugate 
nuclei. This is in agreement with both covariant \cite{ebr20} and Gogny EDF \cite{sch13} calculations in dilute nuclei.

It should be noted that both the small value of the quantality directly obtained from the alpha-alpha potential \cite{zin08} and the present analysis of the dimensionless ratio of the typical alpha-particle size to the internucleon distance, point to
alphas as possible degrees of freedom of a dedicated EFT.

\section{Quantality for spin-orbit effect and shell structure}

\subsection{The key formula}\label{subkey}

Dimensionless quantities are known to be a powerful tool in physics \cite{uza03}.
In addition to quantality, the dimensionless coupling constant is another important one:

\begin{equation}
\alpha\equiv\frac{r_0V_0}{\hbar c}
\label{defcc}
\end{equation}

Eqs (\ref{la}) and (\ref{defcc}) yields the following relation between the quantality and the coupling constant:

\begin{equation}
\eta'\Lambda\alpha^2=1
\label{key}
\end{equation}
 
 where
 
 \begin{equation}
\eta'\equiv\frac{mc^2}{V_0}
\label{etap}
\end{equation}

It should be noted that $\eta$'  is related to velocity effects in the system. For instance, using the T$\simeq$V$_0$ approximation for liquids, where T is the typical kinetic energy 
of the system, leads to, in the non-relativistic case

\begin{equation}
\eta'\simeq 2\left(\frac{c}{v}\right)^2
\label{nrapetap}
\end{equation}

where $v$ is the velocity of the constituents. Therefore, Eq (\ref{key}) highlights the relation between the quantality, the coupling constant, and the velocity of the
system. For instance, it could be used, in an ab-initio or EFT approach, to calculate the quantality from the behavior of the coupling constant and using $\langle$V$\rangle$ instead of V$_0$ as in Eq. (\ref{lagene}):

 \begin{equation}
\Lambda=\frac{\langle V \rangle}{\alpha^2mc^2}
\label{keygen}
\end{equation}

The spin-orbit effect is well-known for being related to the velocity of the constituents, in a quantum framework. Hence, Eq (\ref{key}) could be related to the spin-orbit effect. In a covariant approach based on a non-relativistic reduction of the Dirac equation, it has been shown the spin-orbit effect in many-body systems is driven by the quantity \cite{ebr16}

\begin{equation}
\eta\equiv\frac{mc^2}{V-S}
\label{eta}
\end{equation}

where V and S are the vector and scalar mean-field potentials generated by mediators of the interaction. A large value of $\eta$ (in absolute value) gives a small spin-orbit effect compared to shell effects, whereas a value close to 1 gives a similar magnitude between spin-orbit and shell effects. The sign of $\eta$ gives the energy ordering of the large j and small j states, whose degeneracy has been raised by the spin-orbit effect \cite{ebr16}: $\eta<$0 corresponds to the state with larger j value at larger energy. Typical values of $\eta$ are given in Table I.

In the case of atomic electrons, S=0, and there is only the attractive vector
potential V. For instance, V=-13.6 eV for an electron in the Hydrogen atom. In the case of nuclei, V$\simeq$320 MeV is repulsive, whereas S$\simeq$-400 MeV is attractive. 
In all the cases, the magnitude of the central potential is V$_0$=-(V+S).

Eqs (\ref{etap}) and (\ref{eta}) leads to 

\begin{equation}
\eta=\varepsilon_{VS}\eta'
\label{releta}
\end{equation}
 
 with  
\begin{equation}
\varepsilon_{VS}\equiv\frac{-(V+S)}{V-S}=\frac{V_0}{V-S}
\label{epseta}
\end{equation}

 When there is no scalar mediator, such as in the vast majority of many-body systems, then $\varepsilon_{VS}$=-1. A departure from this value indicates the presence of a scalar mediator in the interaction.  For instance, in the case of nuclei, using the typical values of V and S leads
to $\varepsilon_{VS}$=1/9$\simeq$0.1. $\varepsilon_{VS}$ is hereafter called the Pseudo Spin Symmetry (PSS) breaking coefficient because the $\varepsilon_{VS}$=0 case corresponds to the realization of the PSS \cite{gin97}, namely V=-S. Moreover, the vector and the scalar potentials cannot be of the same sign, because of the repulsive/attractive nature of the spin triplet/singlet dependence of the interactions, implying $\mid\varepsilon_{VS}\mid\le$1. Therefore, the case $\mid\varepsilon_{VS}\mid$=1 corresponds to the maximal PSS violating case.

Using Eqs (\ref{key}) and (\ref{releta}) leads to :

\begin{equation}
\eta\Lambda\alpha^2=\varepsilon_{VS}
\label{key2}
\end{equation}

The relations (\ref{key}) and (\ref{key2}) have a strong physics meaning because they relate quantities involved in various aspects of many-body systems. The interpretation of the coupling constant is obvious. Quantality drives localization aspects (crystal, cluster and QL phases), as discussed above. Finally, $\eta$ can be interpreted, in finite systems such as nuclei or atomic electrons, as the parameter driving the spin-orbit effect \cite{ebr16}. 

\begin{figure}[tb]
\begin{center}
\scalebox{0.70}{\includegraphics{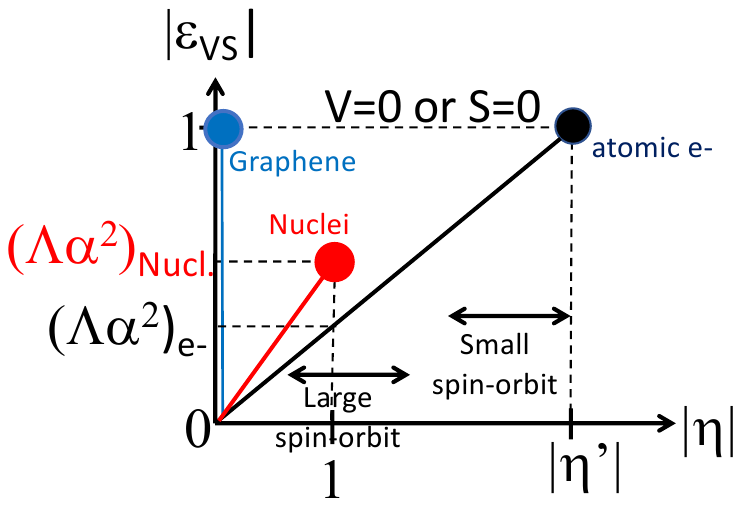}}
 \caption{Sketch of the relation between the PSS coefficient $\varepsilon_{VS}$ and the spin-orbit parameter $\eta$ in the case of 3 bound systems: nuclei, electrons in atoms, and graphene, for a fixed value of V+S$<$0.}    
\label{fig:VS}
\end{center}
\end{figure}

Figure \ref{fig:VS} provides a more detailed overview. For a given fixed central potential value V$_0$=-(V+S), letting V-S vary gives a line in the
 ($\eta$,$\varepsilon_{VS}$) plane, whose slope is $\Lambda\alpha^2$, using Eqs (\ref{releta}),(\ref{key}). 
 In the case of most systems, like electrons in atoms or graphene, there is no scalar mediator in the interaction (S=0). Hence, Eq. (\ref{epseta}) gives $\mid\epsilon_{VS}\mid$=1. In the case of nuclei, there are both vector and scalar mediators, implying $\mid\epsilon_{VS}\mid<$1. Moreover, the value of $\eta$ is close to 1 (see Table I).

Considering the quantality, QL systems have larger slopes than crystal ones. Concerning the interaction effect,
a system at exact pseudo-spin symmetry has V=-S, i.e. V$_0$=$\alpha$=0, and is on the X axis. On the other hand, graphene has an infinite value of $\Lambda$ (see Appendix B), and is on the Y axis.

Let us discuss the above-mentioned dimensionless quantities on nuclei and atomic electrons.
Nuclear states are known to usually behave as quantum liquid states and
therefore $\Lambda \lesssim$ 1, as displayed in Table I. Moreover, the low energy QCD
interaction yields $\alpha\sim$ 1. Eq. (\ref{key}) then provides
$\eta\sim$ 1. This value of $\eta$ allows for strong spin-orbit effect
in finite systems \cite{ebr16}.  The nucleus is therefore a very specific system where
$\eta\sim\alpha\sim\Lambda\sim$1 within about one order of magnitude, as seen in Table I. We define this property as specificness. It should be noted that this definition differs from the naturalness one used in EFT, where renormalized dimensionless low energy constants are $\mathcal{O}(1)$ \cite{kol20}.

In order to understand the origin of this specificness in nuclei, 
the coupling constant Eq. (\ref{defcc}) can be rewritten with the mass of the mediator m$_0$c$^2\equiv\hbar$c/$r_0$:

\begin{equation}
\alpha=\frac{1}{\Lambda}\frac{m_0}{m}
\label{alcc4}
\end{equation}

Therefore, specificness could be interpreted by the close mass of the nucleon and the mesons, namely the 1 GeV scale, themselves related to the scale of chiral symmetry breaking and to the chiral condensate \cite{gel68,iof81}: in nuclei, the mass of the mediator is of the same order of magnitude than
the constituent particle mass (m$_p$/m$_\pi$ $\simeq$7 at most). The m$_{\pi,\sigma}\lesssim $m fact, responsible for $\alpha \sim$ 1, is related to QCD
effects which provide a typical energy scale of few hundreds of MeV. This allows for
similar masses between the lightest baryons and mesons. Eq. (\ref{alcc4})
exhibits the relationship between QCD driven masses ( $\alpha \sim$ 1), the strong spin-orbit effect in nuclei ($\eta \sim$ 1) and their QL behavior ($\Lambda \sim$ 1). In other words, nuclei are specific in the sense that mainly one interaction, namely the strong force, is involved in both their constitution and their interactions.

In the case of electrons in atoms, $\Lambda$ is close to 1 as well (even being close to its maximal value, see section II), which also behave as a QL \cite{mot96}.
 However, this is not a specificness system since the coupling constant is the electromagnetic one: $\alpha$=1/137, yielding $\eta\sim$10$^4$ using Eq.(\ref{key}). Namely, values are far from unity for $\alpha$ and $\eta$. Indeed, the definition of the spin-orbit parameter (Eq. (\ref{eta})) yields
$\eta\sim$m$_e$c$^2$/V$_0\sim$ (10$^5$ eV/10 eV) $\sim$10$^4$, which is in agreement with the experimentally observed magnitude of the spin-orbit effect, compared to the shell one \cite{ebr16}, i.e. the fine structure: 1/$\eta \sim$10$^{-4}$. 

\subsection{Spin-orbit, quantality, and coupling constant analysis of many-body systems }

Let us further investigate the relation (\ref{key2}) between the quantality, the interaction, and the spin-orbit effect, and show how quantality impacts the spin-orbit effect in finite systems. 
The spin-orbit rule \cite{ebr16} is derived from the non-relativistic reduction of the Dirac equation and, therefore, encompasses information on the equation of motion. It gives the magnitude $x$ of the ratio of the HO shell gap to the one induced by the spin-orbit effect, as a function of $\eta$ \cite{ebr16}. 1/$x$ monitors the importance of the spin-orbit effect with respect to the main shell effect: 1/$x$ $\sim$ 1 in nuclei, in agreement with nuclear structure phenomenology showing a spin-orbit degeneracy raising able to build magic gaps. On the contrary, 1/$x$ $\sim$ 10$^{-4}$ for electrons in atoms, in agreement with the hyperfine
structure in atoms.

Using the key relation (\ref{key2}) into the spin-orbit rule \cite{ebr16}, yields:

\begin{equation}
x(\Lambda)=\left |\frac{\epsilon_{VS}}{\Lambda\alpha^2}-1+\frac{\Lambda\alpha^2}{4\epsilon_{VS}}\right |
\label{xalpha}
\end{equation}

where $x$ is the ratio of the HO shell gap to the spin-orbit one.

\begin{figure}[tb]
\begin{center}
\scalebox{0.35}{\includegraphics{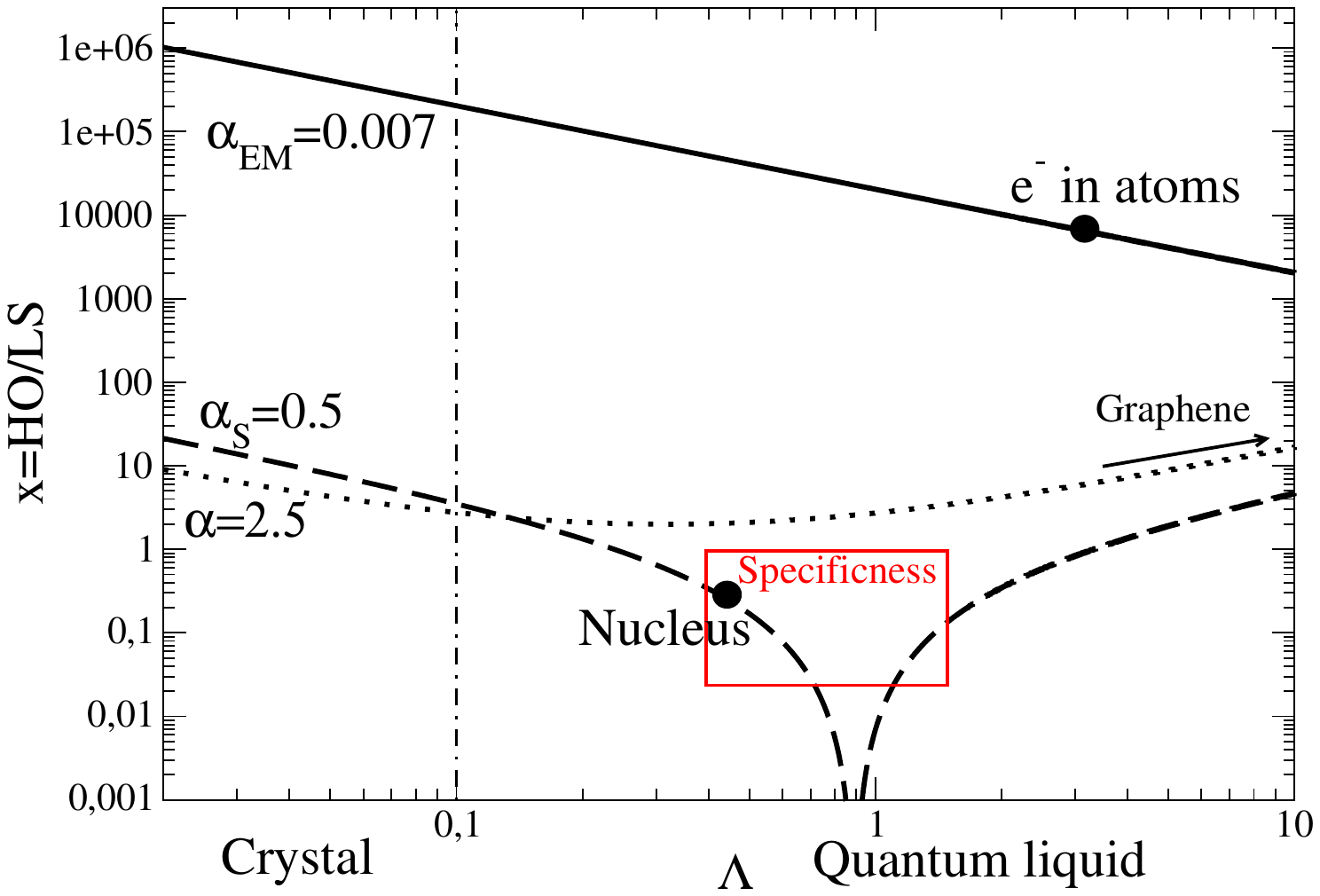}}
 \caption{The relative magnitude of the HO to spin-orbit gaps in a many-body system as a function of the quantality (Eq. (\ref{xalpha})), for various dimensionless coupling constant: the electromagnetic (EM) case (the fine 
 structure constant), the low-energy strong interaction (S) between nucleons, and the case of graphene ($\alpha$=2.5).}    
\label{fig:spcou}
\end{center}
\end{figure}

Fig. \ref{fig:spcou} displays the $x$($\Lambda$) relation, for
both the QL and crystal cases.  In the specific case of a quantum liquid ($\Lambda\simeq $1) 
and in the case of the electromagnetic interaction ($\alpha\simeq$10$^{-2}$,$\epsilon_{VS}$=-1),
Eq. (\ref{xalpha}) reduces to

\begin{equation}
x(\Lambda=1)\simeq\frac{1}{\alpha^2}\simeq 10^4
\end{equation}

recovering the specific fact that the coupling constant is driving the magnitude of the spin-orbit effect of electron in atoms \cite{som}. 
Eq. (\ref{xalpha}) therefore provides a general understanding of the impact of the coupling constant on the
spin-orbit effect, in various systems. Figure \ref{fig:spcou} shows that in a QL ($\Lambda\simeq $1), the large spin-orbit coupling ($x\lesssim$1)
can only be reached in strongly interacting systems whereas the small
spin-orbit coupling ($x>>$1) can only be reached in electromagnetic interacting
systems:  a large spin-orbit effect requires both a coupling constant of the order of unity (such as in the strong
interaction case) and a quantum liquid behavior, as seen on Eq. (\ref{xalpha}). 

In the case of a crystal ($\Lambda\lesssim$ 1), the spin-orbit effect is reduced, by a factor corresponding to the drop of
the quantality value from a QL to a crystal, typically 2-3 orders of
magnitude (Table I). This result is in agreement with dedicated calculations of the spin-orbit effect in crystals \cite{coo78}.

Another interesting case is graphene (see appendix B), where $\eta$=0, $\Lambda$ goes to infinity, and $\alpha$ remains finite at about 2.5.

\subsection{Binding energy vs shell effects dependence on quantality}

In Fermionic many-body systems, it could be relevant to analyze the respective value of the binding energy to the typical energy scale 
at work, namely the major shell gap. A constituent-dominated system means that the binding energy per particle is larger than the typical shell gap, and 
the Fermi energy of the system doesn't change drastically. On the other hand, a shell-structure-dominated system means that the Fermi energy can be small
compared to the shell gap. Let us show how quantality also monitors this relative effect.

In nuclei, the binding energy 
per nucleon is B/A $\sim$ 8 MeV, a bit larger than
the typical magic shell gap which is about 5 MeV between the last level of a shell and the first one of the next shell. 
As a general question, the competition in finite systems between the energy per constituent and the shell effect 
shall be considered. This feature explains why
the addition of one nucleon to the system has a larger effect than magicity effects, 
as seen on the plots of the evolution of the separation energy as a function 
of the neutron or proton number (see e.g. Fig. 1 of Ref.  \cite{lun03}, where the lines usually do not cross each other).

\begin{figure}[]  
      {\includegraphics[width=0.65\textwidth]{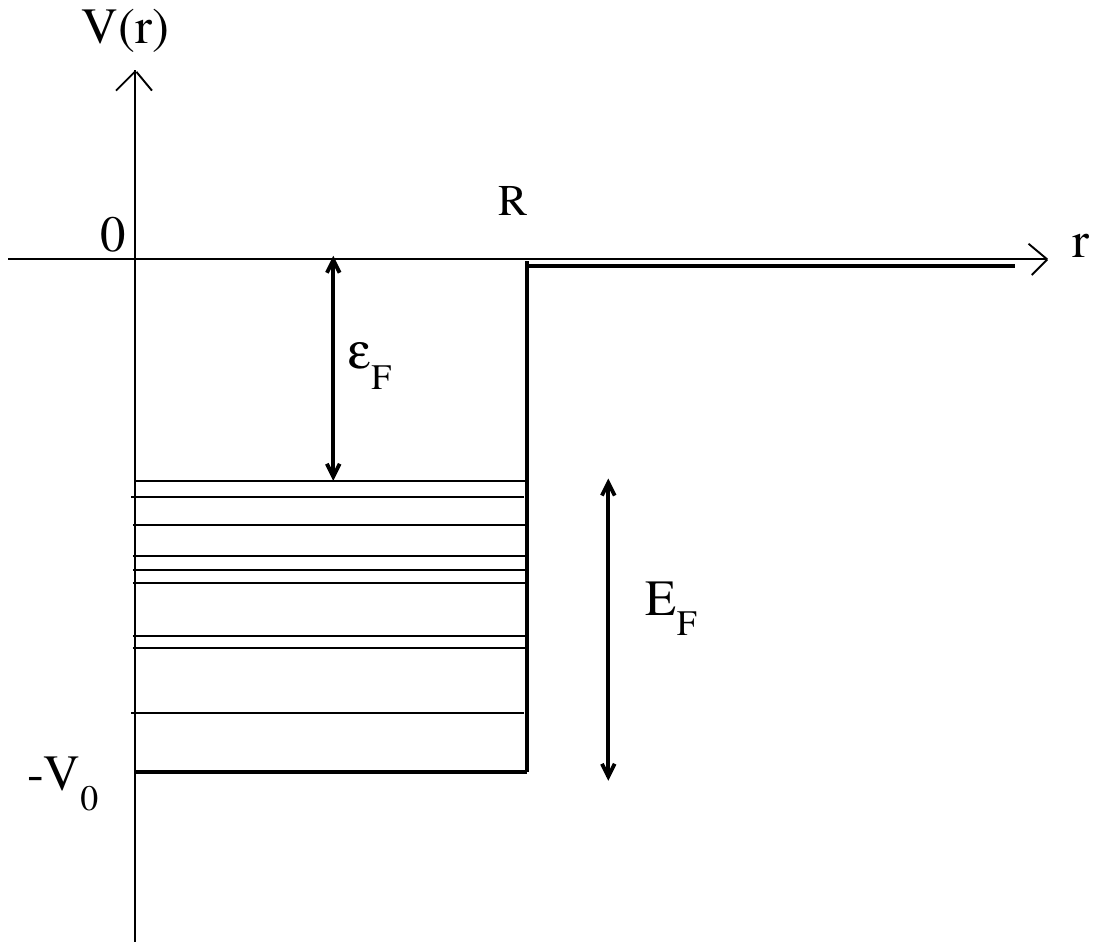}
    }
    \vspace{-1.5cm}
    \caption{The confining potential and relevant quantities}    
    \label{fig:sket}
  \end{figure}

Therefore, the relevant quantity to analyse this behavior is the y ratio defined as

\begin{equation}
\label{y_ratio}
y\equiv\frac{B/A}{\hbar\omega}\simeq \frac{\epsilon_F}{\hbar\omega}
\end{equation}

where $\hbar\omega$ is the typical HO shell gap.
B/A is typically the Fermi energy of the system: Fig.
\ref{fig:sket} displays the quantities to be used with respect to the
confining potential of the system. y $>$ 1 is a constituent-dominated
system where the addition of one constituent can drastically impact the
system: this is the case for nuclei, as known from its phenomenology. y $<$ 1 is a
shell structure dominated system.

The binding energy per nucleon can be approximated by B/A $\sim\epsilon_F=V_0-E_F$ (Fig. \ref{fig:sket}). E$_F$
shall be evaluated from a simple Fermi gas approximation 
which reads \cite{rs80}:

\begin{equation}
\label{ef}
E_F=\frac{\left(9\pi\right)^{2/3}}{8}\frac{\hbar^2}{mr_0^2}(1+\delta)^{2/3}
\end{equation}

where $\delta$=(N-Z)/A is the neutron excess parameter. Over the nuclear chart one safely has 0$\lesssim\delta\lesssim$0.5. 

The typical harmonic oscillator gap is \cite{bm69} 
\begin{equation}
\hbar \omega_0 = {\hbar \over R} \sqrt{ 2 V_0 \over m}, 
\label{eq:hbar}
\end{equation}

It should be noted that the Fermi gas and the HO both correspond to approximate ways of solving the Schr\"odinger equation.

For the numerical evaluation, two points should be considered: i) the Fermi
gas model provides a quantitative value of B/A $\sim$ 8 MeV when V$_0$
$\simeq$ 45 MeV. This rather small value of V$_0$ is due to the approximations of the Fermi gas model. ii)
injecting this value in (\ref{eq:hbar}) gives $\hbar\omega\simeq$10
MeV. The energy gap between the last subshell of a given major shell and the first one of the
next shell (magic gap) is therefore rather of the order of
$\hbar\omega$/2$ \simeq$ 5 MeV. 

This yields for the y ratio discussed above, using Eqs (\ref{ef}) and (\ref{eq:hbar}):

\begin{equation}
y\simeq\frac{V_0-E_F}{\hbar\omega_0/2}=\frac{\sqrt{2}R}{\hbar}\sqrt{mV_0}-
\frac{\left[9\pi(1+\delta)\right]^{2/3}}{4}\frac{\hbar
R}{r_0^2\sqrt{2mV_0}}
\end{equation}

which can be rewritten, with the quantality (\ref{la}):

\begin{equation}
y(\Lambda)\simeq A^{1/3}\left(\frac{2}{\sqrt{\Lambda}}-\frac{\left[9\pi(1+\delta)\right]^{2/3}}{4}\sqrt{\Lambda}\right)
\end{equation}

using R=r$_0$A$^{1/3}$. Considering the quantality value for nucleons ($\Lambda\simeq$0.4, Table I) and the possible range of $\delta$ over the nuclear chart gives

\begin{equation}
y\simeq (1.2;1.7)A^{1/3}
\end{equation}

Therefore, in nuclei y$>1$, meaning that the addition of a nucleon is larger on the binding energy than
the main shell gap.

\section{Generalization of the quantality}

The present study shows the pivotal role of the $\alpha$ (\ref{defcc}), $\eta$ (\ref{eta}) and $\Lambda$ (\ref{la})
dimensionless ratios to characterize several fundamental properties of many-body states. More generally, it is possible
 to build in a systematic way dimensionless
quantities from the 3 basic quantities of the system: V$_0$, r$_0$ and
m. New dimensionless quantities could be produced in the spirit of the quantality using the Pi theorem \cite{buc14}. It should be noted, in a general way, that the value of dimensionless parameters allows the identification of the regime where a given phenomenon (clusterization, quantum liquid behavior, etc.) is dominant and to build an EFT that describes this regime. 
In the following, in order to have only energy units, and for the sake of simplicity, the r$_0$ lengthscale can equivalently be
described by the following mass:

\begin{equation}
m_0c^2 \equiv \frac{\hbar c}{r_0}
\label{yu}
\end{equation}

Setting the V$_0$, m$_0$, and m values imposes two independent ratios built from these.
We choose the one driving the spin-orbit parameter (\ref{etap}):

\begin{equation}
\eta'\equiv\frac{mc^2}{V_0}
\label{etaa}
\end{equation}

and the dimensionless coupling constant:

\begin{equation}
\alpha=\frac{V_0}{m_0c^2}
\label{cca}
\end{equation}

In the case of up to maximum second order for any of
the 3 basic quantities of the system (V$_0$, m$_0$ and m), four independent 
dimensionless quantities can be built. Table \ref{tab:quanta} summarizes the various dimensionless quantities discussed below for several
many-body systems. 

The first quantity 

\begin{equation}
\Lambda\equiv\frac{1}{\eta'\alpha^2}=\frac{m_0^2c^4}{mc^2V_0}
\label{lad1}
\end{equation}

(\ref{lad1}) is the quantality $\Lambda$ and the key formula (see Eq. (\ref{key})). It is related to the action of the system normalized to $\hbar$:

\begin{equation}
{\cal A} \equiv \frac{r_0\sqrt{mV_0}}{\hbar}
\label{action}
\end{equation}

by

\begin{equation}
\frac{1}{{\cal A}^2} = \Lambda
\label{lamact}
\end{equation}

This shows that the action and the quantality of a system provide the same information. Indeed, it is well known that quantum effects in a system are large when its
action is close to $\hbar$ \cite{coh}. This corresponds to
$\cal{A}$$\gtrsim$ 1 and $\Lambda$ $\lesssim$ 1. This is the
quantum liquid case. When quantum effects are smaller, such as in the
crystal case, the action of the system is significantly larger than
$\hbar$: $\cal{A}$$\gg$ 1 and therefore $\Lambda$ $\ll$ 1.

The second quantity which can be built from $\eta'$ and $\alpha$ is

\begin{equation}
ZPEv\equiv\frac{1}{\eta'\alpha}=\frac{m_0}{m}=\sqrt{\frac{2T_0}{mc^2}}\simeq\frac{v_0}{c}\equiv\beta_0
\label{eqzpev}
\end{equation}

It corresponds to the velocity v$_0$ of the constituents, due to the ZPE. Its values in various systems are given in Table I. 
 In the case of nuclei, it has a non-negligible value compared to one. Eq (\ref{eqzpev}) shows that it comes from the specificness of this system. 
Following this interpretation, the velocity of the nucleons is non-negligible compared to the light speed (i.e. at the upper limit of the non-relativistic 
approximation) because the spin-orbit effect is important ($\eta\simeq$1), as well as the magnitude of the interaction ($\alpha\simeq$1).
 
As the third quantity:

\begin{equation}
TI\equiv\frac{\alpha}{\eta'}=\frac{V^2_0}{m_0c^2mc^2}
\label{lad3}
\end{equation}

This quantity is large when the interaction is large and/or the spin-orbit effect is large (small value of $\eta$).
Therefore, it monitors the magnitude of the total interaction, namely the central+spin-orbit effects. Nuclei are expected to
have the largest value of the total interaction, because of the both large spin-orbit and interaction magnitudes. 

The last independent quantity is

\begin{equation}
QV\equiv\frac{1}{\eta'^2\alpha}=\frac{m_0c^2V_0}{m^2c^4}=\frac{\beta_0}{\eta'}
\label{lad2}
\end{equation}

Therefore, QV is the ratio of the zero point motion velocity to the spin-orbit parameter. A large value
 corresponds to a large zero-point motion velocity and a large spin-orbit effect. 
In addition to the ZPE velocity, QV can be also related to the quadratic velocity of the constituents of the system: as an approximation, let us consider 
the kinetic energy of the constituent of the order of V$_0$, as it can be in liquids.
In this case, (\ref{nrapetap}) gives $\eta$'$\simeq$2/$\beta^2$, where $\beta$ is the velocity of the constituent over c. Hence, 

\begin{equation}
QV\simeq \frac{\beta_0\beta^2}{2}
\label{}
\end{equation}

The typical values of the quadratic velocity for various systems are given in Table I.
Once again, in nuclei, this value is the closest to one and the largest, compared to other systems.

\section{Conclusions}

The quantality is a powerful tool allowing to identify various phases in many-body systems. Its upper value corresponds to a minimal radius, which is close to the Bohr radius in the case of atomic electrons, and leads, in the case of nuclei, to the size of the hard-core of the nucleon-nucleon. 
To analyze cluster phases in nuclei, the localization parameter can be related to the quantality. The more general definition of the localization parameter allows to show that both the nucleon number and the low density are control parameters for the occurrence of the cluster phase.

When related to the dimensionless coupling constant, the quantality drives the spin-orbit effect in finite systems. The nucleus is a very specific system: its dimensionless constants are close to the order of the unity.  The nuclear clusterization condition could even be considered as a superspecificness state: all the dimensionless parameters are close to one, including the localization parameter. The impact of quantality on the spin-orbit effect is exhibited, using the spin-orbit rule obtained from a non-relativistic reduction of the Dirac equation. Also, the quantality is shown to drive the binding energy per particle compared to the shell gap.
 
Three additional dimensionless quantities have been discussed. It should be reminded that the quantities used in the present work, contain information from the solved Schr\"odinger equation (of course, in an approximated way), through e.g. the evaluation of the dispersion of the nucleonic wave function obtained by microscopic EDF or HO approximation, or the behavior of the spin-orbit effect through a non-relativistic reduction of the Dirac equation. The present results show how quantality
can be useful to nuclear structure, beyond a mere evaluation of its value in various many-body systems. It can be evaluated in microscopic approaches such as EDF or Ab initio and EFT ones. 

\appendix

\section{The Br\"uckner parameter}

The Br\"uckner parameter, used in condensed matter, can be related to the localization parameter (\ref{locdef2}), in the case of nuclei. 
Behaving quantum-mechanically, the effective strength of the interaction between nucleons, i.e. the extent to which interactions impact the properties of nuclei
and make the latter deviates from the ideal free case, can be measured by a
dimensionless ratio between the mean potential $\braket{V}$ and the mean quantal kinetic $\braket{K}$ energies,
or equivalently the ratio between the mean interparticle distance $\bar{r}$ and 
the Bohr radius $a_B$. This is the so-called
Br\"uckner (also known as the Wigner) parameter (see e.g.~\cite{bon98})
$r_s=\frac{\braket{V}}{\braket{K}}\sim\frac{\bar{r}}{a_B}$. 
In the nuclear case, 
we have
\begin{equation}
r_s=\frac{\hbar\omega}{E_0},
\label{eq:rs}
\end{equation}
where the energy of the harmonic oscillator reads 
$\hbar\omega=\frac{\hbar}{R}\sqrt{\frac{2V_0}{m}}$ and 
$E_0=\frac{\hbar^2}{2m}\left(\frac{3\pi^2\rho}{2}\right)^\frac{2}{3}$
stands for the Fermi energy of nuclear matter at density $\rho$ \cite{rs80}. One gets from Eqs. (\ref{eq:al}) and (\ref{eq:rs}):
\begin{equation}
r_s\simeq\alpha_{loc}^{-2}
\end{equation}
showing that the localisation parameter captures the relevant effects. 

It should be noted that in Plasma physics, these dimensionless parameters are also used to explore the temperature dependence of various phases \cite{bon98}.

\section{Graphene}

Graphene can be described by massless effective charge carriers
using a 2D Dirac equation \cite{nov05}. The effective fine structure constant in graphene is
$\alpha\simeq$2.5 \cite{shy07}. The interaction is of electrodynamics type. Therefore, $\varepsilon_{VS}$=1 because S=0 (see Eq. (\ref{epseta})).
Let us investigate whether the present approach could describe the case of graphene.

In the case of massless constituent (m=0), only two quantities
describe the system: V$_0$ and r$_0$ which are the magnitude and the range
of the effective interaction, respectively. Hence, only one dimensionless
parameter can be built from these two quantities. In the case of
graphene, and following the present lines, it shall be
the coupling constant $\alpha$ (Eq. (\ref{defcc})). 

The quantality (\ref{la}) behaves as $\Lambda\rightarrow\infty$ in the case of graphene
because m=0: graphene could be considered as a
``super''-QL, and  the wave functions of the charge careers shall be
much delocalized. This is in agreement with theoretical studies on
systems obeying to 2D Dirac equations \cite{zie98}.

In the case of graphene, the spin-orbit parameter becomes $\eta$=0 (because m=0), meaning that the spin-orbit effect vanishes ( $\Lambda\rightarrow\infty$ in Eq. (\ref{xalpha})). However, the key relation
(\ref{key}) still holds, i.e. Eq. (\ref{key2}) with $\varepsilon_{VS}$=1:

\begin{equation}
\eta\Lambda\alpha^2=1
\end{equation}

Therefore, graphene case can be described by the present
approach, as a limit case where $\eta$=0, $\Lambda$=$\infty$ and
$\alpha$ remains finite.

\section*{Acknowledgement}
The authors thank M. Chabot, R. Lasseri, A. Mutschler, P. Schuck and D. Vretenar for useful discussions.

\end{document}